\documentclass[twocolumn,aps,prd,epsf]{revtex4}
\usepackage{epsfig}
\usepackage{color}
\usepackage{amsmath}

\begin{document}
\title{ 
Does The Lattice Zero Momentum Gluon Propagator for Pure Gauge SU(3) Yang-Mills Theory Vanish
in the Infinite Volume Limit?
 }
\author{O. Oliveira}
\email{orlando@teor.fis.uc.pt}
\affiliation{CFC, Departamento de F\'{\i}sica, Universidade de Coimbra, Rua Larga, 3004-516 Coimbra, Portugal}
\author{P. J. Silva}
\email{psilva@teor.fis.uc.pt}
\affiliation{CFC, Departamento de F\'{\i}sica, Universidade de Coimbra, Rua Larga, 3004-516 Coimbra, Portugal and \\ School of Physics, The University of Edinburgh, Mayfield Road, Edinburgh EH9 3JZ, United Kingdom}

\begin{abstract}
The Cucchieri-Mendes bounds for the gluon propagator are discussed for the four dimensional pure gauge SU(3) theory. Assuming a pure power law dependence on the inverse of the lattice volume, the lattice data gives a vanishing zero momentum gluon propagator in the infinite volume limit. Our investigation shows that this result is robust against variations of the lattice volumes and corrections to the power law. Moreover, it is consistent with the Gribov-Zwanziger horizon condition and contradicts the SU(2) analysis, which assumed the same type of dependence on the inverse of the volume. Our analysis considers also more general ansatze that, although not conclusive, open the possibility of having $D(0) \ne 0$ in the infinite volume limit. A solution to this puzzle requires further investigations.
\end{abstract}

\maketitle

\section{Introduction and Motivation}

Gluon confinement criteria \cite{AlkGreen} make precise statements about the behavior of
the zero momentum Landau gauge gluon propagator $D(0)$. In particular, the
Gribov-Zwanziger horizon condition \cite{Gribov,Zwan94,Dok,Zwan91a,Zwan91b}
predicts a vanishing propagator at zero momentum. Moreover, in \cite{Zwan91b} it was
shown that, in the minimal Landau gauge, the lattice zero momentum gluon propagator 
vanishes in the infinite volume limit and that same prediction holds for all zero momentum
gluon correlation functions. For the gluon propagator, these predictions are not in line with
early calculations using Dyson-Schwinger equations which required an infinite zero momentum gluon
propagator to explain quark confinement \cite{Bla74,Pa77,Man79},
 i.e. a linearly rising interquark potential.

First principles calculations of the gluon propagator rely either on Schwinger-Dyson equations or
on lattice QCD simulations. 

In what concerns the Schwinger-Dyson equations, recent solutions follow in two distinct classes: 
i) solutions which have a vanishing propagator at $q = 0$ (for a recent review see \cite{Fischer06}); 
ii) solutions with a finite but non vanishing value for the propagator at zero momentum \cite{AgBiPa}.  The reader should be aware that the Schwinger-Dyson equations are an infinite tower of equations
relating the Green's functions of QCD. Finding a solution of the infinite set of equations requires
truncating the tower of equations and parametrizing some of the vertices.

On the other hand, four dimensional lattice simulations for the gluon propagator have been done with
SU(2) and SU(3) gauge theories (see, for example,
\cite{Cu07,Ilg07,Lein98,Fur04,CuccLarge,BerlimLarge,OlSi07,SiOl06,Cuc0712,Stern07,CuOl07}
and references therein). All lattice simulations show a finite $D(0) \ne 0$. 
Furthermore, recent simulations using huge lattice volumes $(27 \mbox{ fm})^4$ for SU(2) 
\cite{CuccLarge} and $(13.2 \mbox{ fm})^4$ for SU(3) \cite{BerlimLarge} give 
a propagator which goes to a constant in the infrared limit. More, the raw lattice
data shows no suppression of $D(q^2)$ in the infrared limit. However, the lattice data shows a
decreasing $D(0)$ as the lattice volume increases and naive extrapolations of $D(0)$ 
to the infinite volume, which typically assume a linear behaviour, always give a finite non vanishing
value around $ 6 - 10$ GeV$^{-2}$. These results are in contradiction with the prediction of a vanishing
$D(0)$ in the infinite volume limit \cite{Zwan91b}. 

This contradiction can be viewed in two ways, either the simulated volumes are still too small 
or the finite volume effects provide sizable corrections to the gluon propagator in the infrared limit
which are not under control. Giving the typical hadronic scales $\sim 1$ fm,
the scale for reflection positivity violation $\sim 1$ fm, and the volumes of the
recent lattice simulations $\sim 13$ fm or larger, it is harder to believe that such volumes are not
sufficiently large.

To overcome the problem of the finite size effects, in  \cite{OlSi07} a method was suggested which
suppresses the finite volume effects. Indeed, the simulations reported in the article use a set of 
different large asymmetric lattices and for all the lattices investigated in \cite{OlSi07}, 
$D(0)$ is finite and non vanishing. 
The reader should be aware that despite the observed differences on
the gluon propagator, clearly due to finite volume and asymmetry effects \cite{AsyEff}, 
the method discussed in
\cite{OlSi07} to access the infrared properties of the gluon propagator give compatible results for 
all lattices investigated in the article. Furthermore, in \cite{OlSi07} no \textit{a priori} assumptions 
about the value of $D(0)$ in the infinite volume limit were made. The new method discussed in that
work provides consistent results for all lattices and points always to a vanishing $D(0)$ in the infinite 
volume limit.

In a previous article \cite{SiOl06} the same authors arrived at a similar conclusion based on 
extrapolations towards the infinite volume of the fitted propagators. 
Presently, \cite{SiOl06,OlSi07} are the only lattice simulations which suggest a result in agreement 
with the Gribov-Zwanziger horizon condition and with the results of \cite{Zwan91b} for the gluon
propagator but they required a closer look at the finite volume effects.

In \cite{Cuc0712}, the authors derived rigorous upper and lower bounds for the zero momentum
gluon propagator of lattice Yang-Mills theories in terms of the average value of the gluon field.
The bounds follow directly from the Monte Carlo approach to lattice simulations. More interesting is the
scaling analysis which the authors have performed for the SU(2) Yang-Mills theory. The scaling
analysis allows to conclude in favor or against a $D(0) = 0$ in the finite volume. For SU(2) the results
show a $D(0)$ going to zero for the two dimensional theory, in agreement with a previous
investigation \cite{Maas}, but a non vanishing $D(0)$ for three and four dimensional formulations.

The aim of this article is to perform a similar investigation for the four dimensional lattice SU(3) 
Yang-Mills theory and, hopefully, to check the infinite volume value for $D(0)$.  To achieve our goal
we use different sets of lattices. The scaling analysis performed to each set of lattices concludes 
always in favor of  a $D(0) \rightarrow 0$ when $V \rightarrow + \infty$, in agreement with Gribov-Zwanziger scenario and \cite{Zwan91b,SiOl06,OlSi07}. 
Note that, although in \cite{Stern07,CuOl07} the SU(2) and SU(3) lattice gluon propagators seem to be 
equal up to momenta below 1 GeV, the results of scaling analysis for the two theories (see \cite{Cuc0712}) suggests
that the volume dependence is not the same for both theories. Our investigation does not provide a
final answer on these questions and further investigations are required to understand if there is a 
difference between the two theories and the reason for such a difference.

The paper is organized as follows. In section \ref{DefLat} we briefly review the Cucchieri-Mendes
bounds and discuss the different sets of lattices used here. Finally, in section \ref{Results} 
we show the results of the bounds and discuss the scaling analysis.

\section{Definitions and Lattice Setup} \label{DefLat}

The Cucchieri-Mendes bounds relate the gluon propagator at zero momentum $D(0)$ with
\begin{equation}
   M(0) ~ = ~ \frac{1}{d \left( N^2_c - 1 \right)} \sum_{\mu, a} \left| A^a_\mu (0) \right| \, ,
\end{equation}   
where $d$ is the number of space-time dimensions, $N_c$ the number of colors,
\begin{equation}
  A^a_\mu (0) ~ = ~ \frac{1}{V} \sum_x A^a_\mu (x)
\end{equation}  
and $A^a_\mu (x)$ is the $a$ color component of the gluon field in the real space. According to
\cite{Cuc0712}, the $D(0)$ is related with $M(0)$ by
\begin{equation}
  \langle M(0) \rangle^2 ~ \le ~ \frac{D(0)}{V} ~ \le d \left(N^2_c - 1\right) \langle M(0)^2 \rangle \, .
  \label{bounds}
\end{equation}  
In the last equation $\langle ~ \rangle$ means Monte Carlo average over gauge configurations. The
bounds in equation (\ref{bounds}) are a direct result of the Monte Carlo approach. 
The interest on these bounds comes from allowing a scaling analysis which can help understanding
the finite volume behaviour of $D(0)$: assuming  that each of the terms in
inequality (\ref{bounds}) scales with the volume according to $ A / V^\alpha$, the simplest possibility 
and the one considered in \cite{Cuc0712}, an $\alpha > 1$  for $ \langle M(0)^2 \rangle$ clearly indicates that $D(0) \rightarrow 0$ as the infinite volume is approached.

In our lattice investigation we used the SU(3) pure gauge Wilson action at $\beta = 6.0$ for a number
of different sets of lattices - see table \ref{lattices}. 
The gauge configurations were generated with version 6 of the MILC code \cite{MILC}.

For the $48^4$ and $64^4$ lattices, the configurations were gauge fixed using an overrelaxation algorithm. For the remaining lattices, the gauge fixing procedure was a Fourier accelerated steepest descent. See \cite{SiOl06} for more details on the definitions of the gluon field and propagator.

\begin{table}
\centering
\caption{Symmetric lattices(left). 
The first column shows the lattice side in fm, the second column reports the
number of lattice points, the third the number of configurations generated and the last column the
first non-vanishing momentum. For the conversion to physical units we used $a^{-1} = 1.94$ GeV
\cite{bali}. Asymmetric lattices (right). The table shows the lattices and the number of configurations for
each lattice.}\label{lattices}
\begin{ruledtabular}
\begin{tabular}{lrrr||rr}
  La         &  $L^4$     & \#          &   $q_{min}$    &  $L^3 \times T$  & \# \\
       (fm)  &                  &  Conf.   &           (MeV)   &                               &   Conf.  \\
\hline
  1.63      &  $16^4$  &         52  &      757     &  $8^3 \times 256$    & 80 \\
  2.04      &  $20^4$  &         72  &      607     &  $10^3 \times 256$  & 80 \\
  2.45      & $24^4$   &         60  &      506     &  $12^3 \times 256$  & 80 \\
  2.86       & $28^4$   &         56  &      434     & $14^3 \times 256$  & 128 \\
  3.26      & $32^4$   &       126  &      380     &  $16^3 \times 256$  & 155  \\
  4.90      & $48^4$   &       104  &     254      &  $18^3 \times 256$  & 150\\
  6.53      & $64^4$   &       120  &     190      &                                      & \\
\end{tabular}
\end{ruledtabular}
\end{table}%

\section{Results} \label{Results}

\begin{figure}
\vspace{0.5cm}
\includegraphics[scale=0.33]{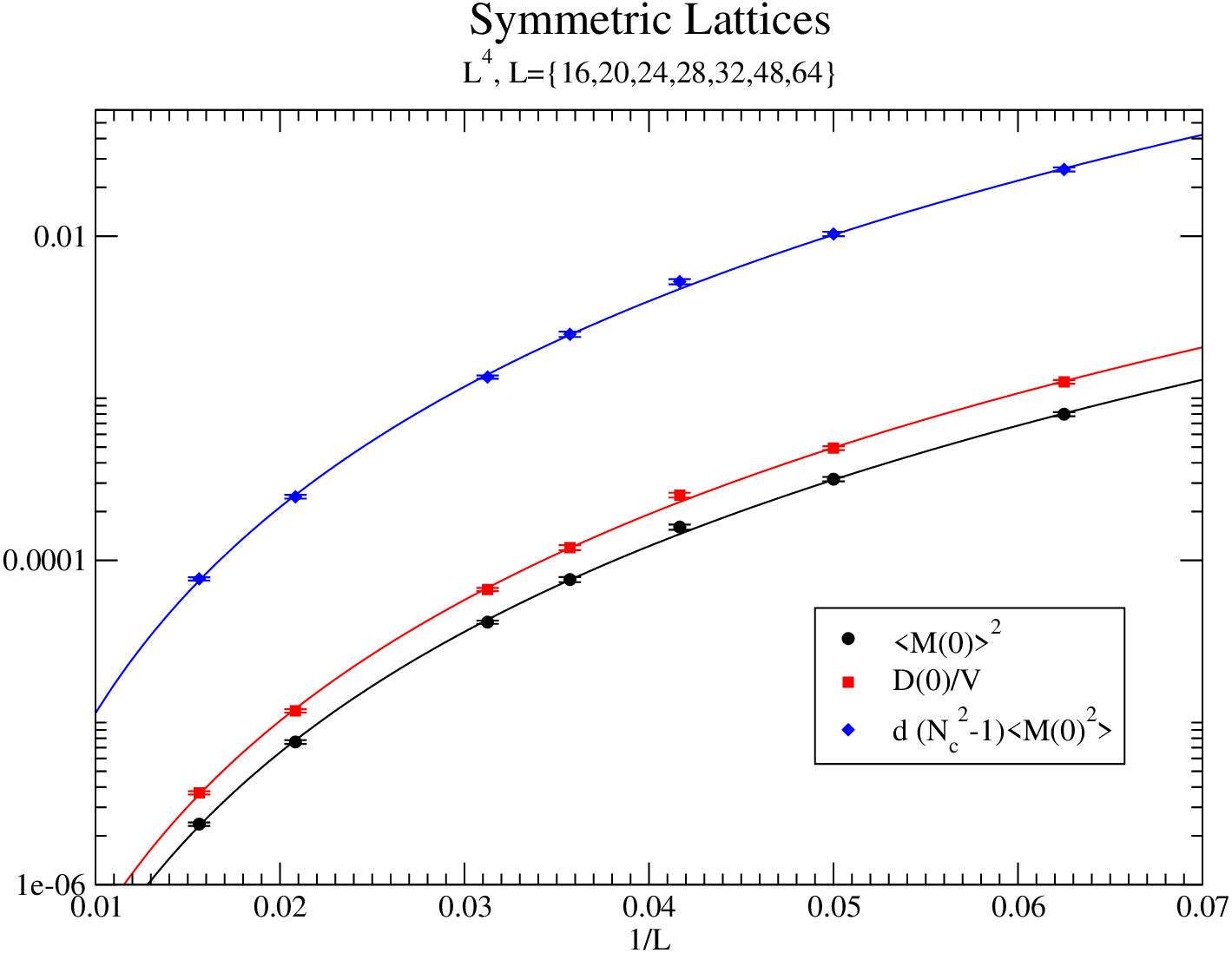}
\caption{Cucchieri-Mendes bounds for  the symmetric set of lattices reported in table \ref{lattices}.}
\label{FigSym}
\end{figure}

\begin{figure}
\includegraphics[scale=0.33]{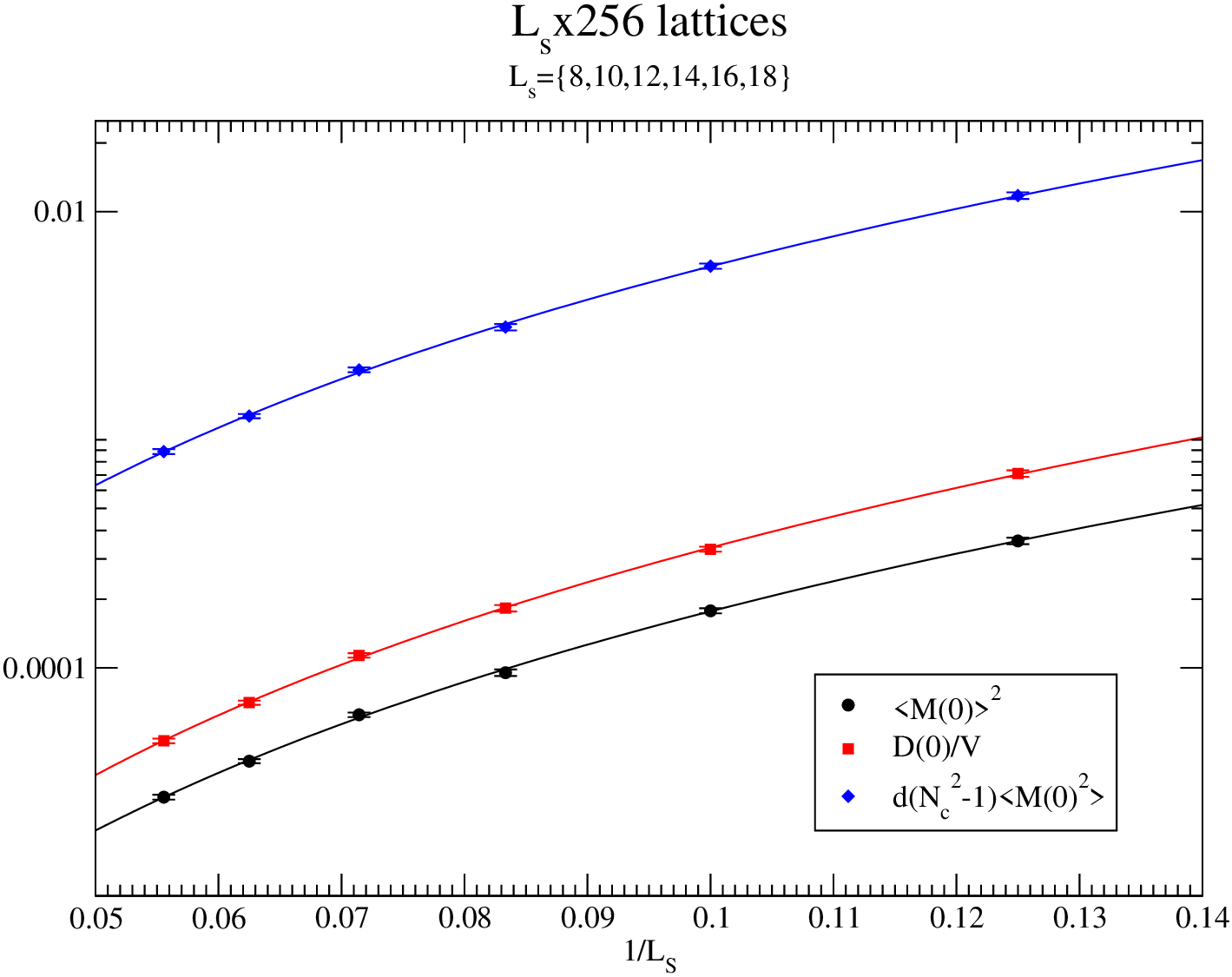}
\caption{The same as figure \ref{FigSym} for the asymmetric lattices reported in table \ref{lattices}.}
\label{FigASym}
\end{figure}

In figures \ref{FigSym} and  \ref{FigASym} we show $\langle M(0) \rangle^2$, $D(0)/V$ and 
$d (N^2_C - 1) \langle M(0)^2 \rangle$ for the symmetric and asymmetric sets of lattices, respectively,
together with the fits to $A / V^\alpha$. As for the SU(2) case \cite{Cuc0712}, the bounds are
verified with $D(0)/V$ being closer to  $d(N^2_c - 1) \langle M(0) \rangle^2$.

The infinite volume limit of $D(0)$ can, in principle, be resolved studying the behaviour of
the upper bound in (\ref{bounds}) with the volume. As stated previously, this was done assuming
that each of the terms in (\ref{bounds}) follows a $A/ V^\alpha$ law. For the symmetric 
lattices in table I, the fits give \footnote{The errors in parameters were computed assuming gaussian error propagation.}
\begin{equation}
\begin{array}{l@{\hspace{0.7cm}}l@{\hspace{0.7cm}}l}
                        & \alpha     & \chi^2_\nu  \\
 \langle M(0) \rangle   & 0.5289(29)  & 2.3 \\
 D(0)/V                 & 1.0587(53)  & 2.4 \\
 \langle M(0)^2 \rangle & 1.0555(57) & 2.4 
 \end{array}
 \nonumber
\end{equation}
where $\chi^2_\nu = \chi^2 /d.o.f.$
Despite the slightly high value for the $\chi^2 / d.o.f.$, the results for $D(0)/V$ and 
$\langle M(0)^2 \rangle$ show that, in the infinite volume limit, $D(0) = 0$. 
Moreover,
they show that the approach to the infinite volume limit is rather slow $ D_V (0) \sim V^{-0.059(5)}$.

A careful looking at figure \ref{FigSym} shows that the $D(0)$ data from the $24^4$ lattice is slightly 
above the fitting curve and is responsible for such high $\chi^2 / d.o.f.$ values. Indeed, removing the
$24^4$ data and repeating the analysis one gets
\begin{equation}
\begin{array}{l@{\hspace{0.7cm}}l@{\hspace{0.7cm}}l}
       & \alpha  & \chi^2_\nu  \\
 \langle M(0) \rangle  & 0.5267(29)   & 1.0 \\
 D(0)/V                          & 1.0558(55)  & 0.8 \\
 \langle M(0)^2 \rangle & 1.0526(59) & 1.0 
 \end{array}
 \nonumber
\end{equation}
i.e. the $\alpha$'s from the two fits are compatible within one standard deviation,  confirming, in this
way, the results of the first fittings. 
The reader should note that in all cases, $\alpha$ from the fits to $D(0)/V$ only becomes compatible
with 1, which would allow a finite $D(0)$ in the limit of infinite volume, within 10 standard deviations.

For the asymmetric lattices, the fits reproduce the same results
\begin{equation}
\begin{array}{l@{\hspace{0.7cm}}l@{\hspace{0.7cm}}l}
       & \alpha  & \chi^2_\nu  \\
 \langle M(0) \rangle  & 0.5320(68)  & 0.8 \\
 D(0)/V                          & 1.104(13)    & 0.5 \\
 \langle M(0)^2 \rangle & 1.062(14) & 0.7 
 \end{array}
 \nonumber
\end{equation}
but with a $\chi^2/d.o.f. \sim 1$. Again, $D(0)$ approaches its infinite volume rather 
slowly $ D_V (0) \sim V^{-0.10(1)}$. 
Note that the $\alpha$'s for the asymmetric set are not compatible, within one standard deviation, with the values computed using the symmetric set. However, to the best of our knownledge there is no reason why the scaling laws for symmetric and asymmetric lattices should be exactly the same. 
Nevertheless, they should provide the same infinite volume limit and this is indeed the case: both sets provide $ D_{\infty} (0) =0 $.

The reader should note that the $\alpha$ from the fits to $D(0)/V$ and $\langle M(0)^2 \rangle$ are above 1 by more than several standard deviations both for the symmetric and asymmetric sets of lattices. In this sense, these results show that $D(0)=0$ in the infinite volume limit\footnote{Besides the lattices reported in table \ref{lattices},
we have performed similar analysis on the following two sets (with $\beta=6.0$): i) $16^3 \times T$ for $T  \in \{16,32,64,128,256\}$; ii) $L^3 \times 32$ for  $L \in \{16,20,24,28,32 \}$. The analysis of these two sets also gives support for $D(0)=0$. Indeed, for the first set we have
 $D(0) \sim V^{-0.042(11)}$ ($\chi^2/d.o.f. = 0.11$) and
$\langle M(0)^2 \rangle \sim V^{-1.090(11)}$ ($\chi^2/d.o.f. = 0.27$). For set ii)
the scaling analysis gives 
$D(0) \sim V^{-0.074(17)}$ ($\chi^2/d.o.f. = 0.67$) and
$\langle M(0)^2 \rangle \sim V^{-1.072(16)}$ ($\chi^2/d.o.f. = 1.68$).}.

The observed slow approach of  $D(0)$ to its infinite volume limit could explain why the previous
naive extrapolations to $V \rightarrow + \infty$ which assumed a linear dependence on $1/V$ 
work so well.

The volumes considered in this work range from  $(1.63 ~\mbox{fm})^4$ up to $(6.53 ~\mbox{fm})^4$,
are smaller than the volumes of the SU(2) analysis \cite{Cuc0712}, where the largest volume is
$\sim (27 ~\mbox{fm})^4$. On the other hand, the SU(2) simulations reported have coarser lattices. 
For our SU(3) simulation $a = 0.102$ fm, to be compared with $a \sim 0.21$ fm for
the SU(2) case; note that both studies use the Wilson action. How the finite volume and finite
lattice spacing effects change the conclusions reported in the two simulations is an open question.

\begin{figure}
\vspace{0.2cm}
\includegraphics[scale=0.33]{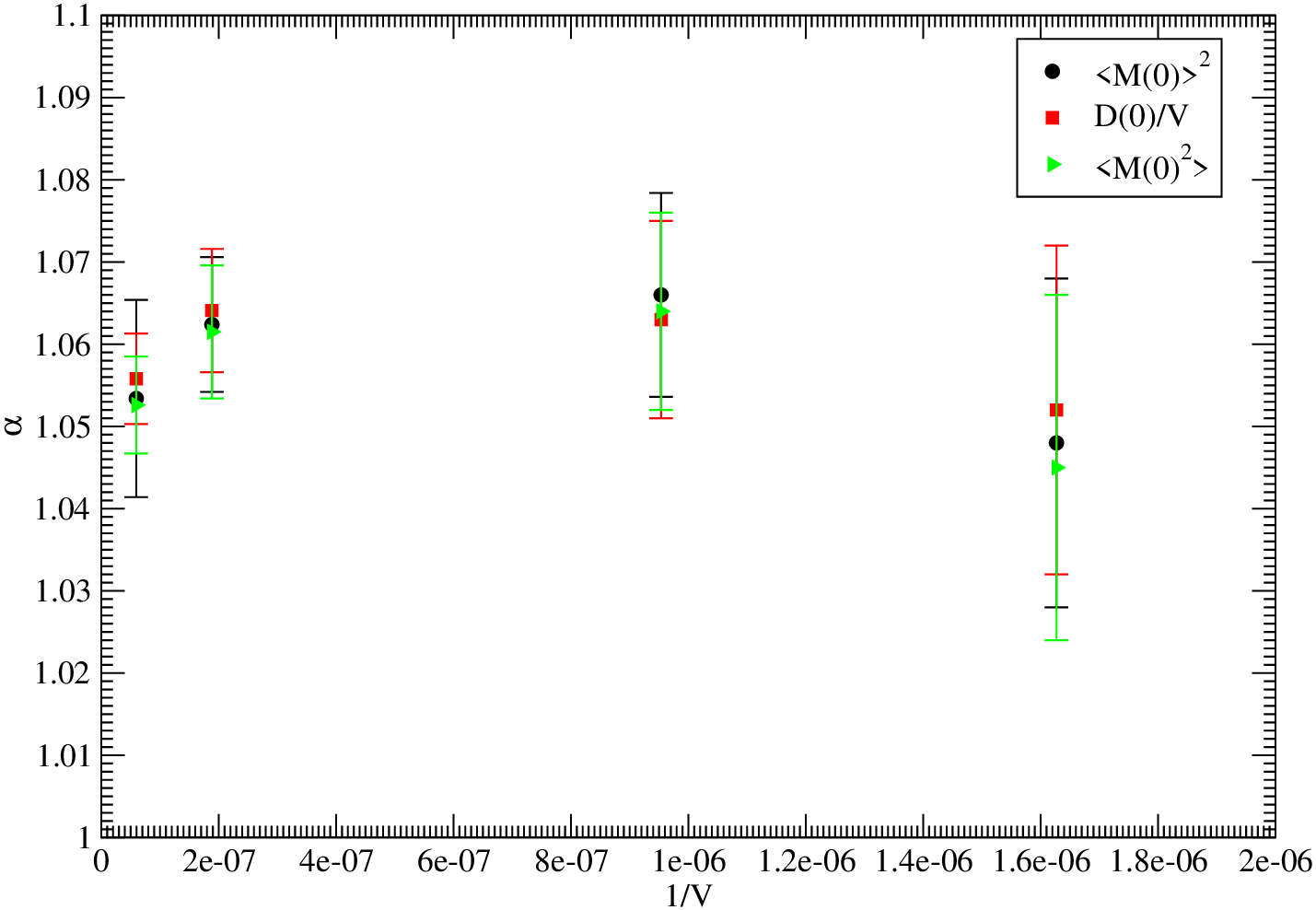}
\caption{$\alpha$ from fitting the lattice data after removing the largest lattices. 
On the x-axis, $1/V$ refers to largest lattice used in the fit, i.e. smallest $1/V$ means including largest
lattices. For $\langle M(0) \rangle$, the figure shows $2 \alpha$.}
\label{alphafig}
\end{figure}

Nevertheless, in order to try to have an estimate of the influence of the lattice volume on our results, in table \ref{fitsvolume} we have performed the same fit to subsets of our symmetric lattices. The $\alpha$ values are, within one standard deviation, stable and compatible with the previous results. In particular, 
one can observe that restricting oneself to volumes up to $28^4$, with a physical volume less than $(3 fm)^4$, one gets $\alpha$ values well above 0.5 or 1  - see also figure \ref{alphafig}.

\begin{table*}[t]
\begin{tabular}{r@{\hspace{0.5cm}}lr@{\hspace{0.8cm}}lr@{\hspace{0.8cm}}lr}
\hline
   &  \multicolumn{2}{c}{$\langle M(0) \rangle$}  & \multicolumn{2}{c}{$D(0)/V$ }     & \multicolumn{2}{c}{$d(N_c^2-1) \langle M(0)^2 \rangle$} \\
\hline
 Lattices          & \multicolumn{1}{r}{$\alpha$} & $\chi^2_\nu$ & \multicolumn{1}{r}{$\alpha$} & $\chi^2_\nu$ & \multicolumn{1}{r}{$\alpha$} & $\chi^2_\nu$ \\
\hline
 16,20,28,32,48,64 & 0.5267(29) & 0.98        &  1.0558(55) & 0.80       &  1.0526(59)  & 1.01    \\
    20,28,32,48,64 & 0.5261(38) & 1.29        &  1.0542(72) & 1.02       &  1.0520(77)  & 1.35    \\
       28,32,48,64 & 0.5216(52) & 1.09        &  1.049(10)  & 1.27       &  1.042(10)   & 1.11    \\
          32,48,64 & 0.5178(60) & 0.58        &  1.043(12)  & 1.41       &  1.035(12)   & 0.56    \\
\hline
\multicolumn{1}{l}{16,20,28}           & 0.524(10)  & 0.14        &  1.052(20)  & 0.0055     &  1.045(21)   & 0.22    \\
\multicolumn{1}{l}{16,20,28,32}        & 0.5330(62) & 0.59        &  1.063(12)  & 0.24       &  1.064(12)   & 0.74    \\
\multicolumn{1}{l}{16,20,28,32,48}     & 0.5312(41) & 0.44        &  1.0641(75) & 0.17       &  1.0615(81)  & 0.53    \\
\multicolumn{1}{l}{16,20,28,32,48,64}  & 0.5267(29) & 0.98        &  1.0558(55) & 0.80       &  1.0526(59)  & 1.01    \\
\hline
\end{tabular}
\caption{Fitting subsets of the symmetric lattices to $A / V^\alpha$, not taking into account the $24^4$. } \label{fitsvolume}
\end{table*}
One could also try to evaluate the contributions from the corrections to the leading behaviour. Having this in mind, the data was fitted to
\begin{equation}
  \frac{A}{V^\alpha} \left( 1 + \frac{\omega}{V} \right) \, .
  \label{fits2}
\end{equation}
The fits to this functional form in (\ref{fits2}) gives
\begin{equation}
\begin{array}{l@{\hspace{0.7cm}}l@{\hspace{0.7cm}}l@{\hspace{0.7cm}}l}
                                              & \alpha  & \omega & \chi^2_\nu  \\
 \langle M(0) \rangle          & 0.5232(53) & 1530 \pm 1954 & 1.1 \\
 D(0)/V                                  & 1.050(10)   & 2518 \pm 3808   & 0.9 \\
 \langle M(0)^2 \rangle      & 1.047(11)   & 2689 \pm 4012 & 1.2 
 \end{array}
 \nonumber
\end{equation}
Note that these new set of $\alpha$'s are, within one standard deviation, compatible with the values
computed assuming a $A/V^\alpha$ behaviour, i.e. the analysis of the corrections to the leading scaling
behaviour do not change the above results. Furthermore, $\omega\sim0$ within the statistical accuracy of the simulation. The fits to the asymmetric lattices data, 
not reported here, lead to the same conclusions.

The above analysis clearly supports a vanishing gluon propagator in the infinite volume, i.e.
$D_\infty (0) = 0$. However, as already discussed in 
\cite{LAT05}, the value of $D_\infty (0)$ depends on the way we model the approach to the infinite volume. The reader should be aware that there is no theoretical guidance about the behaviour of these quantities with the volume. Up to this point, as in \cite{Cuc0712}, we have assumed the simplest behaviour, a power law in the volume. In order to test for the possibility of having a finite and non-vanishing $D_\infty (0)$, one has to consider other fitting functions. In particular we tested the following one,
\begin{equation}
 \frac{C}{V} ~ + ~ \frac{A}{V^\alpha} .
 \label{finitemass}
\end{equation}
A finite and non-vanishing $D_\infty (0)$ requires $C \ne 0$. For the full symmetric set of lattices considered in this work, we get fits with $\chi^2/d.o.f.>2.5$, i.e. the quality of the fits are poor. However, if we do not take into account the $24^4$ data as done previously, the fits give
\begin{equation}
\begin{array}{l@{\hspace{0.5cm}}l@{\hspace{0.5cm}}l@{\hspace{0.5cm}}l@{\hspace{0.5cm}}l}
                                              & A                            & \alpha      & $C$    & \chi^2_\nu  \\
 \langle M(0) \rangle^2     & 208 \pm 228     & 1.21(12)  & 33(7)           & 0.8 \\
 D(0)/V                                  & 263 \pm 247        & 1.18(12)  & 48(14)        & 0.7 \\
 \langle M(0)^2 \rangle      & 6127 \pm 6645   & 1.20(13)  & 1064(236) & 0.9
 \end{array}
 \nonumber
\end{equation}
(note that in the first line we are reporting $ \langle M(0) \rangle^2$ and not $ \langle M(0) \rangle$). 

For the asymmetric set, we get
\begin{tabular}{l@{\hspace{0.2cm}}l@{\hspace{0.4cm}}l@{\hspace{0.4cm}}l@{\hspace{0.4cm}}l}
                                       &            &                      &               &   \\
                                       &  $A$   &  $\alpha$    &  $C$     & $\chi^2_\nu$ \\
  $\langle M(0) \rangle^2$               & $273\pm6020$ & $1.01(31)$  & $-190\pm6102$   &  1.06 \\
  $D(0)/V$                             & $539\pm6849$  & $1.02(35)$    & $-332\pm7151$  & 0.67   \\
  $\langle M(0)^2 \rangle$  & $3957\pm15273$ & $1.14(65)$  & $788\pm2891$   & 0.94  \\
                                       &            &                      &               &   \\
\end{tabular}
So, while for the asymmetric lattices one gets $C=0$, the analysis for the symmetric lattices allows for a finite non-zero $D_\infty (0)$. In this sense, the results of (\ref{finitemass}) are not conclusive.

As before, one could try to check for the stability of the fits considering subsets of symmetric lattices - see table \ref{Cstab}. The results shows that we only get $C\ne0$ when our largest lattice $64^4$ is included in the analysis. All the results of fitting equation (\ref{finitemass}) suggest that either it does not make sense to use (\ref{finitemass}) or larger volumes are required. We would like to recall that a similar analysis assuming $A/V^\alpha$ behaviour does not suffer from the same problem, i.e. the lattice data is well described by $A / V^\alpha$ and the exponent is independent of the subset of lattices used in the analysis.

\begin{table*}[t]
\begin{tabular}{r@{\hspace{0.5cm}}lr@{\hspace{0.9cm}}lr@{\hspace{0.9cm}}lr}
\hline
   &  \multicolumn{2}{c}{$\langle M(0) \rangle^2$}  & \multicolumn{2}{c}{$D(0)/V$ }     & \multicolumn{2}{c}{$d(N_c^2-1) \langle M(0)^2 \rangle$} \\
\hline
 Lattices          & \multicolumn{1}{l}{C} & $\chi^2_\nu$ & \multicolumn{1}{l}{C} & $\chi^2_\nu$ & \multicolumn{1}{l}{C} & $\chi^2_\nu$ \\
\hline
 16,20,28,32,48,64 & $33.0\pm6.7$ & 0.76        &  $48\pm14$     & 0.66      &  $1064\pm236$  & 0.87    \\
    20,28,32,48,64 & $36.7\pm3.4$ & 0.51        &  $53.9\pm8.7$  & 0.76      &  $1203\pm102$  & 0.53    \\
       28,32,48,64 & $36.5\pm3.3$ & 0.99        &  $54.3\pm8.5$  & 1.19      &  $1263\pm40$   & 0.27    \\
\hline
\multicolumn{1}{l}{16,20,28,32}        & $(-1\pm11)\times10^3$ & 0.98  &  $14\pm107$  & 0.50   &  $333\pm2043$   & 1.54    \\
\multicolumn{1}{l}{16,20,28,32,48}     & $8\pm100$ & 0.66         &  $-199\pm3055$ &  0.21   &  $-30\pm5108$  & 0.79    \\
\multicolumn{1}{l}{16,20,28,32,48,64}  & $33.0\pm6.7$ & 0.76      &  $48\pm14$ & 0.66       &  $1064\pm236$  & 0.87    \\
\hline
\end{tabular}
\caption{Stability of C - symmetric lattices without $24^4$ . }\label{Cstab}
\end{table*}

For completeness, we have also considered a generalization of (\ref{finitemass}) and
fitted the lattice data to
\begin{equation}
\omega_1 V^{-\alpha}(1+\omega_2 V^{-\beta})
\label{twoppl}
\end{equation}
(note that equation (\ref{finitemass})
 corresponds to the particular case $\alpha=1$). A full analysis of this ansatz is not easy, given the non-linear nature of the fit. However, the task can be
made easier fixing $\alpha$ to the values obtained from the fits to the pure power law and then
evaluated $\beta$. 
For the symmetric lattices, discarding the data from $24^4$, the results are as follows
\begin{tabular}{l@{\hspace{0.2cm}}l@{\hspace{0.4cm}}l@{\hspace{0.4cm}}l@{\hspace{0.4cm}}l}
                                       &                          &                    &                           &   \\
                                       &  $\omega_1$    &  $\beta$    &  $\omega_2$     & $\chi^2_\nu$ \\
  $\langle M(0) \rangle^2$               &  $93.6\pm1.4$    &  $0.87\pm1.90$  & $240\pm5061$   &   1.24 \\
  $D(0)/V$                             &  $152.2\pm2.1$     &  $0.97\pm2.95$  & $574\pm19718$   &   1.01 \\
  $\langle M(0)^2 \rangle$  &  $83\pm46$   &  $0.79\pm1.89$  & $83\pm1740$  &   1.30 \\
                                       &                          &                    &                           &   \\
\end{tabular}
Note that in all cases and within the present statistical accuracy, $\omega_2=0$.
However, if we fix $\alpha=0.9$, which would imply an infinite $D(0)$, the $\chi^2 / d.o.f.$ are again
around one, 
\begin{tabular}{l@{\hspace{0.3cm}}l@{\hspace{0.4cm}}l@{\hspace{0.4cm}}l@{\hspace{0.4cm}}l}
                                       &                          &                    &                            &   \\
                                       &  $\omega_1$    &  $\beta$    &  $\omega_2$     & $\chi^2_\nu$ \\
  $\langle M(0) \rangle^2$               &  $3.11\pm1.6$    &  0.216(46)  & $50.8\pm8.1$   &   0.67 \\
  $D(0)/V$                             &  $4.2\pm2.6$     &  0.207(43)  & $56\pm15$   &   0.59 \\
  $\langle M(0)^2 \rangle$  &  $98\pm54$   &  0.212(46)  & $50.3\pm9.4$  &   0.78 \\
                                       &                          &                    &                            &   \\
\end{tabular}
i.e. in what concerns the behaviour of D(0) in the infinite volume 
the analysis of fitting (\ref{twoppl}) do not allow to conclude anything. In order to distinguish between 
a $A/V^\alpha$ behaviour or that summarized in equation (\ref{finitemass}) 
simulations with larger volumes are needed and/or simulations with larger number of configurations; unfortunately, both solutions are beyond our current computational power. 
Of course, a theoretical guidance on the volume behaviour of these quantities would be also welcome.

\section{Conclusions}

In this paper the Cucchieri-Mendes bounds for SU(3) Yang-Mills theory
are investigated and, like for the original analysis
of the SU(2) gauge propagator \cite{Cuc0712}, we start assuming that
the lattice data has a $V^{-\alpha}$ volume dependence. This
functional form describes quite well the lattice $D(0)$ both for the
SU(2) and SU(3) propagators. For
SU(3), the data supports the prediction of \cite{Zwan91b}, i.e.  $D_
\infty (0) = 0$ in minimal Landau gauge, and the Gribov-Zwanziger
confinement scenario. This is a major difference between the SU(2) and
SU(3) simulations. What is the origin for such difference is an open
question which deserves further investigations.
We would like to recall the reader that, in this work, a stability
analysis was performed and the results look rather robust against a
change in the volumes used in the fits.
Furthermore, the scaling analysis is consistent with the results
reported in \cite{OlSi07}, where the authors have shown
that the infrared gluon propagator is compatible with a pure power law
$(p^2)^{2\kappa}$, with $\kappa\sim0.53$. Recall
that this previous result implies, again, a $D_{\infty}(0)=0$.

Besides the scaling analysis assuming a $V^{-\alpha}$ behaviour, more
general ansatze for the dependence with the
lattice volume were considered. The results of using these ansatze are
not conclusive but open the possibility of having a $D_\infty (0) \ne
0$. In this sense, the simulations discussed in this article do not
provide a definitive answer about
$D_\infty (0)$. Clearly, a solution to the puzzle require SU(3)
simulations with larger volumes and a deeper theoretical understanding
of the behaviour of $D(0)$ and $M(0)$ with the lattice volume.

One issue which was not considered neither here nor in the SU(2)
simulation \cite{Cuc0712},
is the effect due to the Gribov copies
\cite{Maas08,Cu97,Silva04,Silva07}. However, given that the $\alpha$
values reported here are several $\sigma$'s above the critical value
of 0.5 for $\langle M(0) \rangle$ and 1 for $D(0) /V$ and $\langle
M(0)^2 \rangle$, it seems unlikely that Gribov copies would change our
conclusion. 

\acknowledgments

Part of the present work was funded by the FCT grant  SFRH/BPD/40998/2007 and
projects POCI/FP/81396/2007 and POCI/FP/81933/2007.
Simulations have been done on the supercomputer Milipeia at the University of Coimbra.


\end{document}